\documentclass{jfm}

\usepackage{graphicx}
\usepackage{natbib}
\usepackage{xcolor}
\ifCUPmtlplainloaded \else
 \checkfont{eurm10}
 \iffontfound
 \IfFileExists{upmath.sty}
 {\typeout{^^JFound AMS Euler Roman fonts on the system,
 using the 'upmath' package.^^J}%
 \usepackage{upmath}}
 {\typeout{^^JFound AMS Euler Roman fonts on the system, but you
 dont seem to have the}%
 \typeout{'upmath' package installed. JFM.cls can take advantage
 of these fonts,^^Jif you use 'upmath' package.^^J}%
 }
 \else
 \fi
\fi

\ifCUPmtlplainloaded \else
 \checkfont{msam10}
 \iffontfound
 \IfFileExists{amssymb.sty}
 {\typeout{^^JFound AMS Symbol fonts on the system, using the
 'amssymb' package.^^J}%
 \usepackage{amssymb}%
 \let\le=\leqslant 
 \let\ge=\geqslant 
 }{}
 \fi
\fi

\ifCUPmtlplainloaded \else
 \IfFileExists{amsbsy.sty}
 {\typeout{^^JFound the 'amsbsy' package on the system, using it.^^J}%
 \usepackage{amsbsy}}
 {}
\fi

\newcommand\Pran{\mbox{\textit{Pr}}} 
\newcommand\Ray{\mbox{\textit{Ra}}} 
\newcommand\Nus{\mbox{\textit{Nu}}} 
\newcommand\kin{\mbox{\textit{E}}_\mathrm{kin}} 
\newsavebox{\astrutbox}
\sbox{\astrutbox}{\rule[-5pt]{0pt}{20pt}}

\newcommand\thalf{\ensuremath{{\textstyle\frac{1}{2}}}}

\newcommand\pdiff[2]{\frac{\partial#1}{\partial#2}}
\newcommand\pddiff[2]{\frac{\partial^2 #1}{\partial#2^2}}

\title[Kinetic Energy Transport in Convection]{Kinetic Energy Transport in Rayleigh--B\'{e}nard Convection}

\author[K. Petschel, S. Stellmach, M. Wilczek, J. L\"ulff and U. Hansen]
{K. Petschel$^1$
 \thanks{Email address for correspondence: klaus.petschel@uni-muenster.de},\ns
S. Stellmach$^1$
M. Wilczek$^2$\break
J. L\"ulff$^3$ and U. Hansen$^1$}

\affiliation{$^1$Institut f\"ur Geophysik, Westf\"alische Wilhelms-Universit\"at, M\"unster, D-48149, Germany\\[\affilskip]
$^2$Department of Mechanical Engineering, The Johns Hopkins University, Baltimore, USA\\[\affilskip]
$^3$Institut f\"ur Theoretische Physik, Westf\"alische Wilhelms-Universit\"at, M\"unster, D-48149, Germany
}

\pubyear{2010}
\volume{650}
\pagerange{119--126}
\date{?; revised ?; accepted ?. - To be entered by editorial office}
\begin{document}

\maketitle
\begin{abstract}
The kinetic energy balance in Rayleigh--B\'{e}nard convection is investigated for the Prandtl number range $0.01\le\Pran\le 150$ and for fixed Rayleigh number $\Ray=5\cdot10^{6}$.
The kinetic energy balance is divided into a dissipation, a production and a flux term. 
We discuss profiles of all terms and find that the different contributions to the energy balance can be spatially separated into regions where kinetic energy is produced and where kinetic energy is dissipated.
Analysing the Prandtl number dependence of the kinetic energy balance, we show that the height-dependence of the mean viscous dissipation is closely related to the flux of kinetic energy.
We show that the flux of kinetic energy can be divided into four additive contributions, each representing a different elementary physical process (advection, buoyancy, normal viscous stresses and viscous shear stresses). 
The behaviour of these individual flux contributions is found to be surprisingly rich and exhibits a pronounced Prandtl number dependence. Different flux contributions dominate the kinetic energy transport at different depth, such that a comprehensive discussion requires a decomposition of the domain into a considerable number of sub-layers. On a less detailed level, our results reveal that advective kinetic energy fluxes play a key role in balancing the near-wall dissipation at low Prandtl number, whereas normal viscous stresses are particularly important at high Prandtl number. Finally, our work reveals that classical velocity boundary layers are deeply connected to the kinetic energy transport, but fail to correctly represent regions of enhanced viscous dissipation.
\end{abstract}
\section{Introduction\label{sec: introduction}}
Convection is ubiquitous in nature as many flows in geo- or astrophysics are driven by buoyancy.
Prominent examples are convection in the Earth's mantle, the Earth's core, the interior of giant planets or the outer layer of the sun. 
Convection also plays a major role in many technical and engineering applications, and has therefore been the subject of a large number of studies \citep{Ahlers2009,Chilla2012}. 
The simplest model system for the investigation of buoyancy-driven flows is the Rayleigh--B\'{e}nard configuration, a thermally driven flow within a plane fluid layer heated from below and cooled from above. 

A recurring theme in both theoretical descriptions and phenomenological pictures of Rayleigh--B\'{e}nard convection is the partition of the fluid layer into distinct regions, such as boundary layers close to the wall and a turbulent bulk region in the center \citep{Grossmann2000,Ahlers2009}, sometimes with a further mixing zone embedded in between the two \citep{Castaing1989}. 
These layers are typically defined in terms of experimentally measurable quantities, such as vertical profiles of temperature and velocity, which often show markedly different behaviour in the different regions \citep{Tilgner1993,Du_Puits2007,Zhou2011}. 
From a theoretical point of view however, it is the relative strength of dissipation within the layers that is thought to chiefly control the dynamics of convection, and in particular the macroscopic heat transport \citep{Grossmann2000}. 
Unfortunately, direct experimental measurements of dissipation are still beyond current experimental techniques \citep{Wallace2009}. 
The connection between the experimentally determined layers and the dissipation processes is thus not straight forward, and requires additional phenomenological modelling, such as the popular assumption that the viscous dissipation is mainly generated by shear within a Prandtl--Blasius type layer \citep{Grossmann2000,Ahlers2009,Chilla2012}. 
Surprisingly little work has been performed on testing the validity of these phenomenological models and their underlying assumptions. 
The situation is further complicated by the fact that different definitions of the spatial extent of the layers are used in the literature \citep{Kerr2000,Lam2002,Breuer2004,Zhou2010,Li2012,Shi2012,Scheel2012,Wagner2012}, resulting in different quantitative layer thicknesses strongly depending on the exact definition used. 

A first step towards a  better understanding of the spatial distribution of viscous dissipation in Rayleigh--B\'{e}nard convection has recently been made by \citet{Petschel2013}, using highly resolved numerical simulations which immediately provide local dissipation rates. 
The authors show that layer definitions based directly on dissipation disagree with more classical boundary layer definitions, and are indeed more suitable to explain the heat transfer observed in their simulations. 
Building up on these results, \citet{Scheel2014} have shown recently that classical boundary layers and dissipation layers exhibit similar Rayleigh number scalings up to $\Ray\approx 10^{7}$, whereas they exhibit opposite scalings for higher Rayleigh numbers.
The authors hypothesise whether this change in the trend with $\Ray$ can be related to the transition from soft to hard convective turbulence.
Interestingly, this transition cannot be observed for classical velocity boundary layers \citep{Scheel2012}.

A thorough understanding of dissipation in Rayleigh--B\'{e}nard convection however goes beyond a pure identification of its spatial distribution.
The physical processes responsible for the spatially inhomogeneous dissipation need to be understood in detail. 
While global energy conservation requires average dissipative losses to be compensated by corresponding energy sources, both processes often occur in a spatially separated manner.
For example, kinetic energy is always produced by buoyancy forces in the bulk, whereas the dissipation is often dominated by deformation work close to the boundaries.
A full picture thus needs to include the transport from the source to the dissipation regions, which can be realised by many different physical processes.
Again, a very limited number of studies have targeted these questions, with notable examples by \citet{Deardorff1967} and \citet{Kerr2001}.

In this paper, we extend the existing results by providing a detailed picture of the kinetic energy generation, transport and dissipation in Rayleigh--B\'{e}nard convection.  
A systematic parameter study for varying Prandtl numbers is presented. 
We show that various classical boundary layer definitions can be interpreted directly in terms of kinetic energy fluxes. 
The kinetic energy transport from the bulk to the dissipation dominated near-wall region is analysed in detail, and is shown to be accomplished by processes which strongly depend on the Prandtl number. 

The paper is organised as follows. In section \ref{sec: model}, we discuss our approach in detail. 
In particular, we show that the pressure term in the kinetic energy transport equation, which hampered the analysis in previous works \citep{Kerr2001}, can be decomposed into several parts, each of which is readily interpreted in terms of elementary physical processes. 
The spatial distribution of kinetic energy sources, sinks and fluxes is then computed from direct numerical simulations, and the results are presented in section \ref{sec: profiles}--\ref{sec: deviations}. 
This detailed analysis reveals which physical processes dominate at a given depth and provides a consistent picture of the generation of kinetic energy, its journey towards the boundaries and its final dissipation. 
The dynamics is found to be surprisingly rich, requiring a decomposition of the domain into a considerable number of sublayers, each characterised by different physical processes at work. 
Detailed conclusions are presented in section \ref{sec: discussion}.
\section{Method \label{sec: method}}
\subsection{Governing equations}
The Rayleigh--B\'{e}nard problem is governed by the non-dimensional Bous\-sinesq equations
\begin{eqnarray}
\label{eqn:momentum}
     \pdiff{u_i}{t}+u_j\pdiff{u_i}{x_j}&=& \pdiff{\tau_{ij}}{x_j}+\Pran\Ray\,\theta\delta_{i3}\,,\\
 \label{eqn:heat}
     \pdiff{\theta}{t}+u_j\pdiff{\theta}{x_j}&=& \pddiff{\theta}{x_j}+u_{3}\,,\\
 \label{eqn:continuity}
     \pdiff{u_i}{x_i}&=&0\,,
\end{eqnarray}
where $u_i$ is the velocity component, $\theta$ the deviation from the conductive temperature profile and $\tau_{ij}=-P\,\delta_{ij}+2\,\Pran\,e_{ij}$ is the stress-tensor for incompressible Newtonian fluids.
Here, $ e_{ij}=\thalf\left(\pdiff{u_i}{x_j}+\pdiff{u_j}{x_i}\right)$ is the rate-of-strain tensor and $P$ is the kinematic pressure.

The equations have been made dimensionless by using the system height $L$ as a length scale and the thermal diffusion time $L^2/\kappa$ as a time scale, where $\kappa$ denotes the thermal diffusivity. The system is then characterised by two non-dimensional control parameters only, the Rayleigh number and the Prandtl number:\begin{equation} \label{eqn:ray_pran}
     \Ray=\frac{\alpha g \Delta T L^3}{\nu\kappa}\,,\quad\Pran=\frac{\nu}{\kappa}
\end{equation}
Here, $\alpha$ denotes the thermal expansion coefficient, $g$ the acceleration of gravity, $\Delta T$ the temperature difference between the bottom and the top and $\nu$ denotes the kinematic viscosity.
The top and bottom boundaries are impermeable and kept at a fixed temperature.
No-slip boundary conditions are used for the velocity field.
\subsection{Numerical model\label{numerics}}
To investigate the properties of the kinetic energy transport in detail, the time evolution of the velocity, pressure and temperature fields is studied by means of three-dimensional direct numerical simulations.
For a fixed Rayleigh number $\Ray=5 \times 10^6$, the Prandtl number range $0.01 \le \Pran \le 150$ is systematically explored.
The simulations in a cubic Cartesian box with periodic boundary conditions in the horizontal directions are carried out using an accurate pseudo-spectral method. 
Fourier series are used for spatial discretisation in horizontal direction, whereas Chebyshev basis functions are used in vertical direction.
Temporal discretisation is accomplished through a second-order, semi-implicit Adams--Bashforth time stepping with backward differentiation.
For further details on the numerical method, see \citet{Stellmach2008}.
Since Chebyshev basis functions naturally provide high resolution near the boundaries, the bottleneck of resolving the local kinetic energy balance is a sufficiently high resolved bulk region.
Even for this moderate Rayleigh number, spatial resolutions up to $576^3$ grid points were necessary to adequately resolve the Kolmogorov scales within the bulk at low $\Pran$ \citep{Shishkina2010, Grotzbach1983}.
\section{Kinetic energy balance and decomposition of the flux\label{sec: model}}
In this section we introduce the well-known kinetic energy equation for Rayleigh--B\'{e}nard convection and review the different contributions to the kinetic energy balance.
We will first consider the effects of kinetic energy production, dissipation and fluxes.
To obtain further insights we then decompose the fluxes into several contributions, each of which is caused by a different physical effect.

\subsection{Kinetic energy balance \label{sec: kin energy bal}}
By multiplying the Navier--Stokes equation (\ref{eqn:momentum}) with the velocity $u_i$ and using incompressibility (\ref{eqn:continuity}) we obtain the kinetic energy equation
\begin{equation}\label{eqn:kin_energy}
     \pdiff{}{t}\kin+\pdiff{}{x_i}\left(u_i \kin-u_j\tau_{ij}\right)= -2\Pran\,e_{ij}^2+\Pran\Ray\,\theta u_3\,,
\end{equation}
where $\kin=\thalf u_i^2$ denotes the kinetic energy.
This equation represents a budget equation of the form \citep{Deardorff1967,  Kerr2001}
\begin{equation} \label{eqn:balance}
     \pdiff{}{t}\kin+\pdiff{}{x_i}J_{i}=-\epsilon+S \, ,
\end{equation}
where the right-hand side comprises a sink $\epsilon$ and a source $S$.
The left-hand side includes a temporal change of the kinetic energy as well as a flux term, which spatially redistributes kinetic energy.
By comparing (\ref{eqn:kin_energy}) with (\ref{eqn:balance}) we have
\begin{equation} \label{eqn:source}
     S=\Pran\Ray\,\theta u_3\,,
\end{equation}
which means that the kinetic energy source, which comes from the work done by buoyancy forces, is proportional to the advective heat transport.
The sink is identified as
 \begin{equation}
      \epsilon=2\Pran\,e_{ij}^2\,.
 \end{equation}
All processes which do not change the net budget, but spatially redistribute kinetic energy, can be expressed as the divergence of the total kinetic energy flux
\begin{equation} \label{eqn: kinetic energy flux}
     J_{i}=u_i\kin-u_j\tau_{ij}\,.
\end{equation}
This means that advection and stress transport kinetic energy from regions where net energy is produced to regions where net energy is dissipated.
\subsection{Flux decomposition\label{sec: decompose flux}}
The flux can be further decomposed, and we start by considering the pressure contributions.
A major difficulty in understanding the kinetic energy flux (\ref{eqn: kinetic energy flux}) is the non-locality of the pressure, which is governed by the elliptic equation
\begin{equation}\label{eqn:pressure}
     \pddiff{P}{x_i}=-\pdiff{u_j}{x_i}\pdiff{u_i}{x_j}+\Pran\Ray\,\pdiff{\theta}{x_3}
\end{equation}
that follows from taking the divergence of the momentum equation (\ref{eqn:momentum}).
We can split up the pressure into three contributions, one due to buoyancy,
\begin{equation}\label{eq:templabel37} 
     \pddiff{P}{x_i}^{(\mathrm{buo})} =\Pran\Ray\,\pdiff{\theta}{x_3}\,\quad\mbox{with\ }\quad \left. \pdiff{P}{x_3}^{(\mathrm{buo})}\right|_{x_3=0,1}=0\,,\phantom{\Pran\,\pddiff{u_3}{x_3}}
\end{equation}
and another one due to advection,
\begin{equation}\label{eq:templabel38} 
     \pddiff{P}{x_i}^{(\mathrm{adv})}  =-\pdiff{u_j}{x_i}\pdiff{u_i}{x_j}\,\quad\mbox{with\ }\quad \left. \pdiff{P}{x_3}^{(\mathrm{adv})} \right|_{x_3=0,1}=0\,.\phantom{\Pran\,\pddiff{u_3}{x_3}}
\end{equation}
Finally, the pressure induced by viscous stresses at the boundary is accounted for by a Laplace equation,
\begin{equation}
     \pddiff{P}{x_i}^{(\mathrm{vis})}  =0\phantom{\pdiff{u_j}{x_i}\pdiff{u_i}{x_j}}\,\quad\mbox{with\ }\,\,\quad \left. \pdiff{P}{x_3}^{(\mathrm{vis})}\right|_{x_3=0,1}=\Pran\,\pddiff{u_3}{x_3}\,.\phantom{0}
\end{equation}
The stress tensor can thus be written as
\begin{equation}
     \tau_{ij}=-\left(P^{(\mathrm{adv})} +P^{(\mathrm{buo})} +P^{(\mathrm{vis}) }\right)\,\delta_{ij}+2\,\Pran\,e_{ij}\,.
\end{equation}
Using this expression, the kinetic energy flux can be expressed as
\begin{equation} 
     J_i= u_i\kin+u_j\left[\left( P^{(\mathrm{adv})}+ P^{(\mathrm{buo})}+ P^{(\mathrm{vis})}\right)\,\delta_{ij}-2\,\Pran\,e_{ij}\right] \,.
\end{equation}
The overall kinetic energy flux is hence composed of the flux caused by advective processes, the flux due to buoyancy pressure and of the flux due to viscous processes.
The viscous flux can be further separated into fluxes involving normal and shear stresses.
Here, viscous normal stresses are connected to the diagonal of the rate-of-strain tensor, $\mathrm{diag}(e_{11},e_{22},e_{33})_{ij}$, while viscous shear stresses are connected to the off-diagonal elements of the rate-of-strain tensor $e_{ij}-\mathrm{diag}(e_{11},e_{22},e_{33})_{ij}$.
The individual flux contributions are then defined as:
\begin{eqnarray}
\label{eqn: advective flux}
     &J^{(\mathrm{adv})}_{i}&=u_i\kin+ u_i\,P^{(\mathrm{adv})}\,,\\
\label{eqn: buoyancy flux}
     &J^{(\mathrm{buo})}_{i}&=u_i\,P^{(\mathrm{buo})}\,,\\
\label{eqn: viscous normal flux}
     &J^{(\mathrm{norm})}_{i}&=-2\,\Pran\,u_j\,\mathrm{diag}(e_{11},e_{22},e_{33})_{ij} + u_i\,P^{(\mathrm{vis})}\,,\\
\label{eqn: viscous shear flux} 
     &J^{(\mathrm{shear})}_{i}&=-2\,\Pran\,u_j\left(e_{ij}-\mathrm{diag}(e_{11},e_{22},e_{33})_{ij}\right)\,.
\end{eqnarray}
In summary, the kinetic energy balance equation (\ref{eqn:balance}) can be rewritten as
\begin{equation} \label{eqn:decomposed balance}
     \pdiff{}{t}\kin+\pdiff{}{x_i}\left(J^{(\mathrm{adv})}_{i}+J^{(\mathrm{buo})}_{i}+J^{(\mathrm{norm})}_{i}+J^{(\mathrm{shear})}_{i}\right)=-\epsilon+S\,,
\end{equation}
i.e., the kinetic energy budget consists of a production term, a dissipation term and of four spatial redistribution terms.

\subsection{Horizontally and temporally averaged kinetic energy balance \label{sec: av kin energy bal}}
In the following we will analyse the horizontally and temporally averaged kinetic energy balance.
These averages are denoted by the following nomenclature: $\langle\ldots\rangle_v$ represents an average over the entire volume and over time, and $\langle\ldots\rangle$ represents an average over horizontal planes and time.
In the statistically stationarity case the kinetic energy budget, for this horizontally homogeneous system, simplifies to
\begin{equation} \label{eqn:av_balance} 
     \pdiff{}{x_3}\left\langle J_{3}\right\rangle=-\left\langle \epsilon\right\rangle+\left\langle S\right\rangle\, ,
\end{equation}
where all quantities depend on height only.
The averaged individual fluxes can be simplified to:
\begin{eqnarray}
\label{eqn:av advective flux}
     &\left\langle J^{(\mathrm{adv})}_{3}\right\rangle&=\left\langle u_3\kin\right\rangle+ \left\langle u_3\,P^{(\mathrm{adv})}\right\rangle\,,\\
\label{eqn:av buoyancy flux}
     &\left\langle J^{(\mathrm{buo})}_{3}\right\rangle&=\left\langle u_3\,P^{(\mathrm{buo})}\right\rangle\,,\\
\label{eqn:av viscous normal flux}
     &\left\langle J^{(\mathrm{norm})}_{3}\right\rangle&=-\,\Pran\,\pdiff{}{x_3}\left\langle u_3^2\right\rangle +\left\langle u_3\,P^{(\mathrm{vis})}\right\rangle\,,\\
\label{eqn:av viscous shear flux} 
     &\left\langle J^{(\mathrm{shear})}_{3}\right\rangle&=-\,\Pran\, \pdiff{}{x_3}\left\langle \kin\right\rangle 
\end{eqnarray}
In the following the term ``profile'' will be used to refer to these height-dependent horizontally and temporally averaged quantities.
Because of the up-down-symmetry of the Rayleigh--B\'{e}nard configuration, the profiles are symmetric about mid height, such that only the lower half of the domain will be shown for all profiles.
Similarly, all layers that we are going to introduce on the basis of these profiles, will be defined in reference to the bottom wall.

To ensure comparability between different Prandtl numbers, all profiles are normalised by the volume averaged viscous dissipation rate, which is linked to $\Nus$ by the relation:
\begin{equation} \label{eqn:e_u} 
     \left\langle\epsilon_u\right\rangle_v=\Pran\,\Ray \left(\Nus-1\right)\,,
\end{equation}
where the Nusselt number is defined as the sum of advective and conductive heat transport, \begin{equation}\label{eqn:nusselt}
     \Nus-1=\left\langle \theta u_3\right\rangle-\pdiff{}{x_3}\left\langle \theta\right\rangle\,.
\end{equation}
Since the total heat flux through any plane must be constant at every height, the horizontally and temporally averaged Nusselt number equals its volume average at every height.
\section{Kinetic energy balance profiles\label{sec: profiles}}
In this section we use direct numerical simulations to explicitly compute the individual contributions to the kinetic energy balance (\ref{eqn:av_balance})--(\ref{eqn:av viscous shear flux}).
We first focus our discussion on the source, the sink and the total flux.
After that we turn to a detailed discussion of the individual flux contributions (\ref{eqn:av advective flux})--(\ref{eqn:av viscous shear flux}).

\subsection{Source \label{sec: source profile}}
\begin{figure}
      \centering \includegraphics{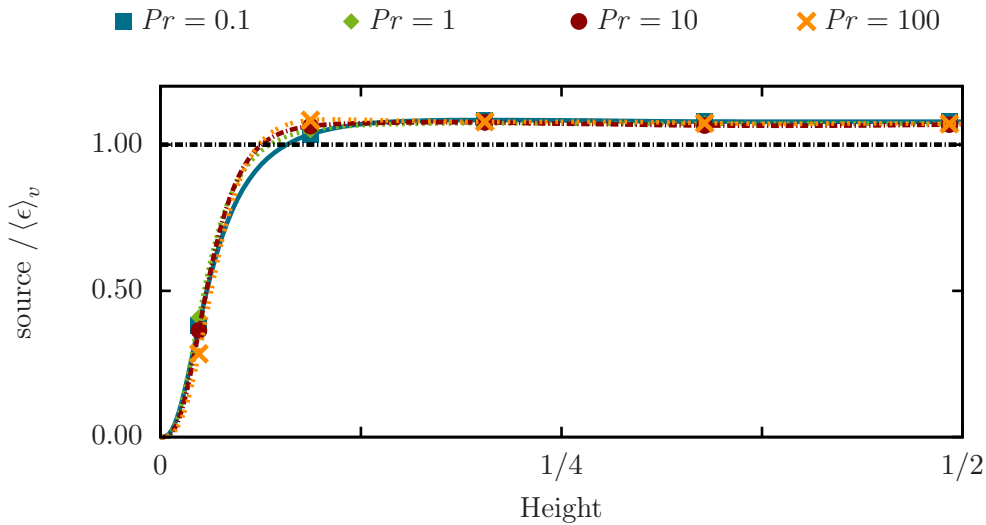}
     \caption{\label{fig: source}Profiles of the kinetic energy source for different Prandtl numbers. }
\end{figure}

The production profile, $\Ray\Pran\left\langle \theta u_{3}\right\rangle$, is displayed in figure~\ref{fig: source} for different Prandtl numbers.
It is proportional to the advective heat transport, $\left\langle\theta u_{3}\right\rangle$.
Since the averaged total heat flux through any plane must be constant at every height, the kinetic energy production is also directly connected to the conductive heat transport: 
\begin{equation} \label{eqn:hor_av_source}
     \frac{\left\langle S\right\rangle(x_{3})}{\Pran\,\Ray}=(\Nus-1)+\pdiff{}{x_{3}}\left\langle\theta\right\rangle(x_{3})\,. 
\end{equation}
The kinetic energy source profile can thus be calculated from the temperature profile and the Nusselt number.

Figure~\ref{fig: source} shows the kinetic energy source profile for different Prandtl numbers.
The profiles are almost independent of the Prandtl number.
All cases exhibit two distinct regions: One associated with an isothermal, well-mixed bulk in the centre of the domain, and a second one related to a near-wall layer, which is characterised by strong temperature gradients.
The bulk region exhibits an almost constant kinetic energy production, while the kinetic energy production in the near-wall layer rapidly decays towards the wall.
Nearly the entire temperature drop occurs within this near-wall region.
Because of the impermeable walls ($u_3|_{x_3=0,1}=0$), the kinetic energy source must be zero at the boundaries.
Both regions can be naturally separated by the crossing of the source profiles with the volume-averaged dissipation (\ref{eqn:e_u}) (which equals the volume-averaged source).
By relation (\ref{eqn:hor_av_source}), this is also the vertical position of the maximum of $\left|\left\langle \theta \right\rangle \right|$, which is sometimes used to define the thermal boundary layer.
Since $\partial/\partial_{x_3} \left\langle\theta\right\rangle \le 1$ in the centre of the domain for a completely mixed fluid, relation (\ref{eqn:hor_av_source}) also reveals the Nusselt number as an upper bound for the source profile in the bulk.
\subsection{Sink}
\begin{figure}
      \centering\includegraphics{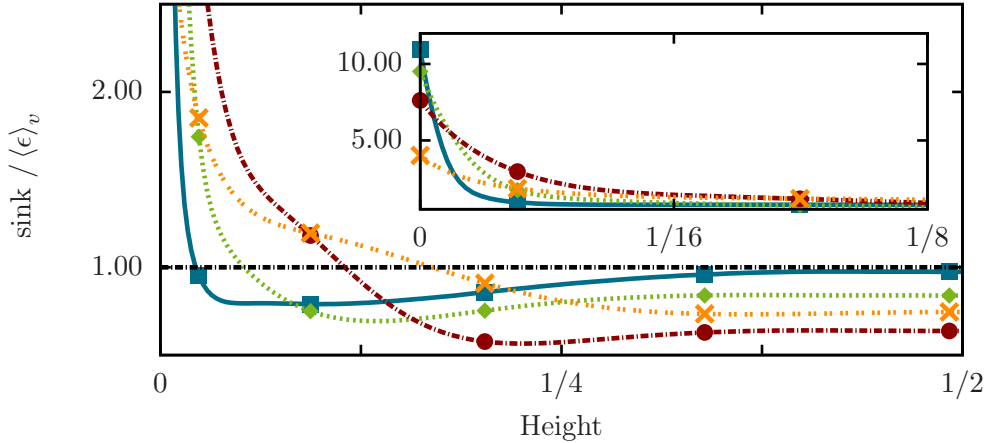}
     \caption{\label{fig: sink} Profiles of the kinetic energy sink, i.e. the viscous dissipation rate for different Prandtl numbers. The inlet of the graph shows a boundary layer blow up to display the respective ``wall dissipation''.}
\end{figure}
The sink in the kinetic energy budget is given by the viscous dissipation term.
Its profile is shown in figure~\ref{fig: sink} for different Prandtl numbers.
Similar to the kinetic energy source, the sink profile can be separated into two regions: A bulk region, characterised by low dissipation rates, and the dissipation layer introduced by \citet{Petschel2013} with significantly enhanced dissipation close to the boundary.
The edge of the dissipation layer is defined as the vertical position at which the dissipation rate equals its volume-averaged value.

In contrast to the kinetic energy production, the transition region in figure~\ref{fig: sink} distinctively varies with the Prandtl number.
For high Prandtl numbers the viscous dissipation increases monotonically towards the boundaries and smears out the transition of bulk and dissipation layer.
The opposite is observed for low Prandtl numbers where the transition is much sharper.

In the centre of the domain the viscous dissipation rate profile is constant for all Prandtl numbers.
However, the amplitude strongly depends on $\Pran$.
For $\Pran = 0.1$ the bulk dissipation rate almost equals its volume average, whereas for Prandtl numbers $\Pran = 1$ and $\Pran = 10$ the bulk dissipation rate is much smaller.
For even larger Prandtl numbers ($\Pran \ge 20$) this tendency reverses, and the viscous dissipation rate in the centre of the domain begins to increase again, as illustrated by the case $\Pran = 100$ in figure~\ref{fig: sink}.

In the near-wall region the kinetic dissipation profile increases significantly towards the boundaries for all Prandtl numbers as can be seen in the inset of figure~\ref{fig: sink}.
The dissipation immediate at the wall, $\epsilon|_{x_3=0,1}$, decreases for increasing Prandtl numbers.
Since the kinetic energy production vanishes at the boundary, the wall dissipation must be balanced entirely by the kinetic energy fluxes.
 
\subsection{Flux} 
\begin{figure}
      \centering\includegraphics{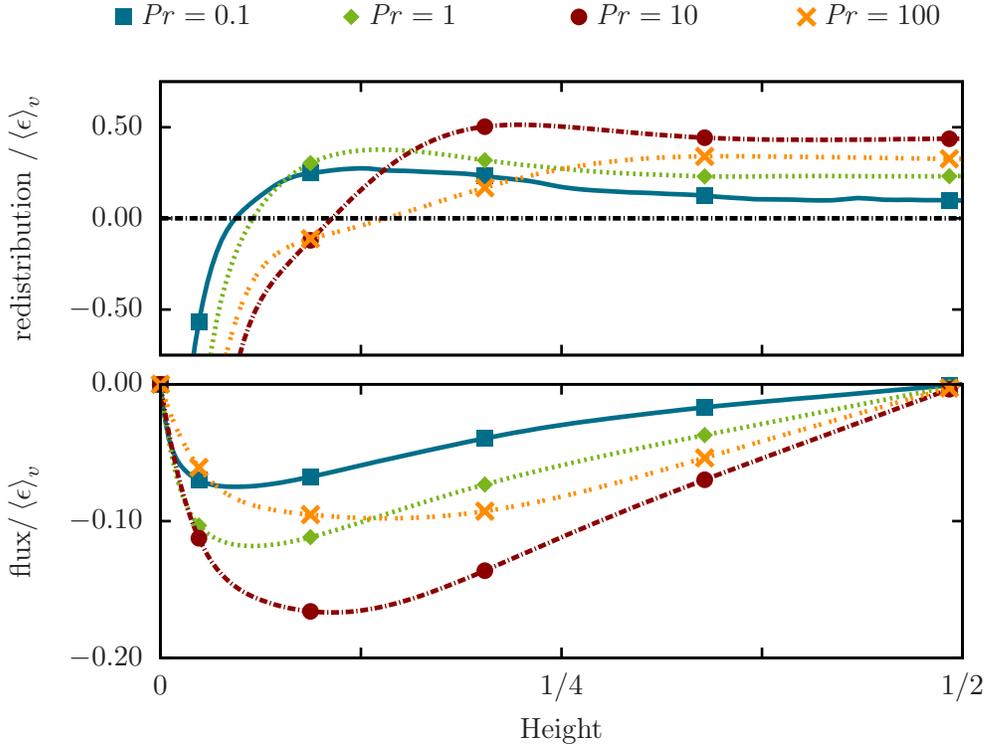}
     \caption{\label{fig: flux}
			Profiles of the kinetic energy redistribution, i.e. the divergence of flux, (upper graph) and the vertical kinetic energy flux (lower graph) for different Prandtl numbers.}
\end{figure}

In addition to the production rate and the viscous dissipation rate, the redistribution of kinetic energy, i.e. the divergence of kinetic energy flux, completes the kinetic energy budget.
The profile of the total flux divergence and the total flux profile are shown in figure~\ref{fig: flux} for different Prandtl numbers.
A negative flux indicates a transport of kinetic energy towards the boundaries whereas a positive flux indicates a transport towards the centre. 
As can be inferred from the figure, the system transports kinetic energy from the centre towards the walls in the entire domain.
Since the Rayleigh--B\'{e}nard configuration is a closed system with up-down-symmetry, all kinetic energy flux profiles have to vanish at the boundaries and at the centre of the domain.

A negative divergence of the flux characterises deposition of kinetic energy whereas a positive flux divergence signifies a removal of surplus kinetic energy.
Because the flux divergence is equal to the difference of kinetic energy source and viscous dissipation rate (cf.~(\ref{eqn:balance})), it exhibits negative values near the boundaries and positive values in the bulk as can be seen in the upper graph of figure~\ref{fig: flux}.
These regions are separated by the zero-crossing of the divergence of the flux, corresponding to the maximum of the kinetic energy flux.
The distance between the boundary of the domain and the vertical position of this maximum increases for increasing Prandtl number.

Since the kinetic energy source does not strongly depend on the Prandtl number, the Prandtl dependence of the viscous dissipation must be linked to that of the flux.
Accordingly, several characteristics of the kinetic energy sink profile in figure~\ref{fig: sink} can also be observed in the flux profile.
Similar to the sink profile, the flux profile also strongly depends on $\Pran$: For high Prandtl number the flux profiles are much flatter than for low Prandtl number and do not exhibit a distinct maximum.
With decreasing $\Pran$ this maximum becomes more pronounced, which indicates that for decreasing Prandtl numbers the boundary between the bulk and the near-wall region becomes sharper.
Again, the Prandtl number tendency of the flux reverses for very high $\Pran$.
The maximum of the flux increases with increasing Prandtl numbers $\le10$ and decreases again for very high Prandtl numbers.
This indicates a regime transition between $\Pran=10$ and $\Pran=100$.

To characterise the Prandtl number dependence of the kinetic energy balance away from the boundaries, it is instructive to investigate the profiles of the individual transport mechanisms. 
The kinetic energy flux consists of four partial fluxes as has been shown in section~\ref{sec: decompose flux}.
The profiles of the individual flux divergences along with their sum is shown in figure~\ref{fig: flux-all} for high and low Prandtl numbers, which will be discussed in the following sections.

\subsubsection{Kinetic energy transport by the individual fluxes for low Prandtl numbers\label{individual low Pr}}
\begin{figure}\centering
     \includegraphics{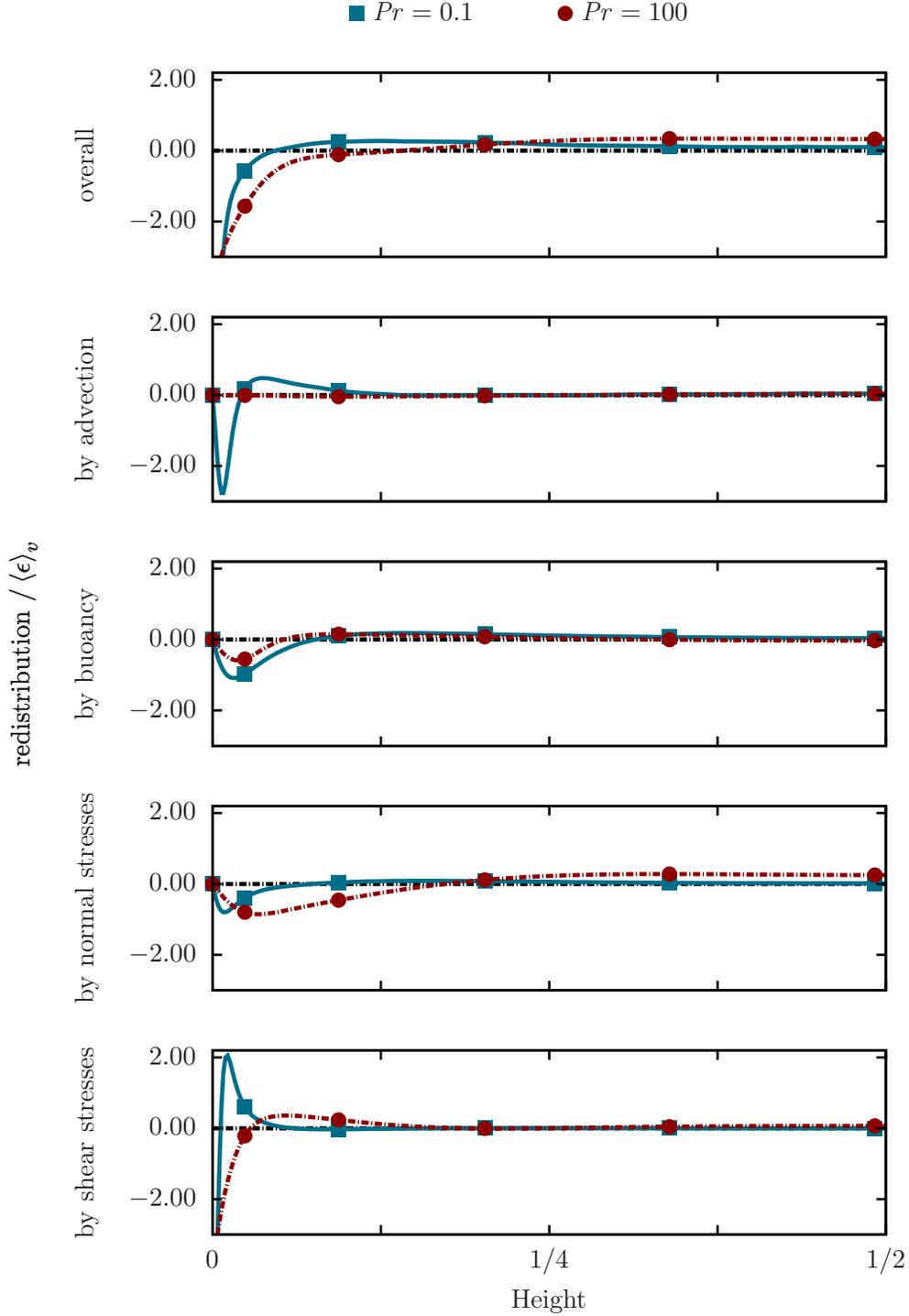}
     \caption{\label{fig: flux-all}
Redistribution of kinetic energy given by the profiles of the divergence of fluxes for high (blue solid line) and low (red dashed line) Prandtl numbers. The uppermost graph shows the redistributions by the sum of all individual fluxes, whereas the bottom four graphs show the redistribution by individual fluxes (cf. section~\ref{sec: decompose flux}).}
\end{figure}

We first focus on our results at $\Pran = 0.1$.
As we have discussed in the previous sections, the kinetic energy is mainly produced in the bulk region and mainly dissipated in the near-wall region.
The total flux redistributes the kinetic energy from the bulk into the near-wall region.
We will discuss in the following the individual redistributions for $\Pran=0.1$ starting from the centre of the domain and proceeding towards the boundaries.
As we have seen above, in the central region of the domain the redistribution of kinetic energy decreases with decreasing Prandtl numbers.
The small redistribution for low Prandtl numbers shows that there is only little surplus of kinetic energy in this region.
As a consequence, the production is almost balanced by the dissipation.
This can be explained by an increasingly statistically homogeneous flow in the bulk region with decreasing $\Pran$.

Figure~\ref{fig: flux-all} shows that the remaining surplus of kinetic energy in the centre is transported towards the walls mostly by buoyancy-pressure flux and normal-stress flux.
Both fluxes deposit the removed kinetic energy in the near-wall region.
The advective flux does not influence the centre, but rather strongly redistributes kinetic energy closer to the walls.
It dominates the transport in the near-wall region between the turbulent bulk and the thin boundary layer close to the rigid wall.

As discussed earlier, the flux divergence must balance the dissipation at the boundaries.
An analytical evaluation of the individual flux contributions, exploiting the no-slip boundary conditions, yields further insights.
In fact, it can be shown that all flux divergences vanish at the boundary, except for the divergence of viscous shear flux.
We can conclude that the ``wall dissipation'' is balanced by the divergence of the shear-stress flux only:
\begin{equation} \label{eqn:wall dissipation}
     \left.\vphantom{\pdiff{}{x_3}\left\langle J^{(\mathrm{shear})}_{3}\right\rangle}\left\langle\epsilon\right\rangle\;\right|_{x_3=0,1}=\left.\pdiff{}{x_3}\left\langle J^{(\mathrm{shear})}_{3}\right\rangle\right|_{x_3=0,1}=\left.-\Pran\left\langle\left(\pdiff{}{x_3}\left( u_2+u_1\right)\right)^2\right\rangle\right|_{x_3=0,1} \, .
\end{equation}
Thus, the redistribution by shear stress deposits kinetic energy much closer to the wall than all other fluxes and is particularly important in the very-near-wall region.
It may thus be regarded as a boundary-driven process.

On the basis of the individual flux profiles in figure~\ref{fig: flux-all}, it can be concluded that the vertical structure of the redistributions can be separated into three regions. 
In the central region only little kinetic energy is transported towards the walls, reflecting a tendency towards homogeneous flow at low $\Pran$.
Closer to the wall, kinetic energy is transported by the advective flux from the bulk towards the boundary.
In the direct vicinity of the wall, the shear-stress flux redistributes kinetic energy to the boundary, where the strongest dissipation takes place.

Interestingly, our results show that the influence of the boundaries extends beyond the small viscous region that is central to many theories of convective heat transport. Beyond this region, the walls induce flow inhomogeneities that cause a strong advective kinetic energy flux. Accounting for this effect is crucial for any comprehensive view of kinetic energy transport in low Prandtl number convection. 

\subsubsection{Kinetic energy transport by the individual fluxes for high Prandtl numbers\label{individual high Pr}}
Similar to the low-$\Pran$ case, the individual redistributions for high Prandtl numbers will be discussed from the centre to the boundary.
For $\Pran = 100$ the production in the bulk region is significantly higher than the dissipation. 
This leads to enhanced vertical redistribution of kinetic energy within the bulk.
Figure~\ref{fig: flux-all} shows that the kinetic energy is transported almost exclusively by normal-stress flux from the centre towards the boundary.
All other fluxes approximately vanish in the bulk.
The vertical position of the minimum of the divergence of the normal-stress flux, i.e. the position of maximal deposition of kinetic energy, is close to the boundary. 
Thus normal-stress flux transports kinetic energy from the centre through the entire domain into the very-near-wall region.

Closer to the wall, buoyancy pressure also starts to redistribute kinetic energy, but in a limited region.
Redistribution by advection is negligible in the entire domain.
Similar to the low-$\Pran$ number case, shear-stress flux redistributes kinetic energy from the very-near-wall region towards the boundary.
Although the shear-stress redistribution profile maintains its general shape across Prandtl numbers, the region which is dominated by shear-stress flux expands with increasing Prandtl number.

The profiles of the individual redistributions suggest that the kinetic energy transport is dominated by viscous effects for high Prandtl numbers.
In contrast to the crucial role of advective flux in the low $\Pran$ case, the normal-stress flux transports kinetic energy from the centre directly to the very-near-wall region.
The transport of kinetic energy from the centre all the way into the boundary region may be seen as an indicator for flow features extending throughout the whole domain, as observed for high Prandtl numbers in previous studies \citep{Breuer2004,Kerr2000}.
\section{Dissipation-, Source- and Flux-Layers\label{sec: layers}}
As we have seen in the previous section, the profiles vary considerably as a function of height.
This allows to separate the domain into distinct regions.
We now introduce the concept of layers for the individual contributions as a quantitative measure to distinguish these regions, generalising the dissipation layers introduced by \citet{Petschel2013}.
We will then analyse the Prandtl number dependence of the layers related to the source, sink and flux.
After that we will turn to a detailed discussion of the layers associated to the individual flux contributions (\ref{eqn:av advective flux})--(\ref{eqn:av viscous shear flux}). 
We close the section by a comparing with classical velocity boundary layers.

\subsection{Definition of the Layers\label{sec: layer thickness}}
\begin{figure}
     \centering \includegraphics{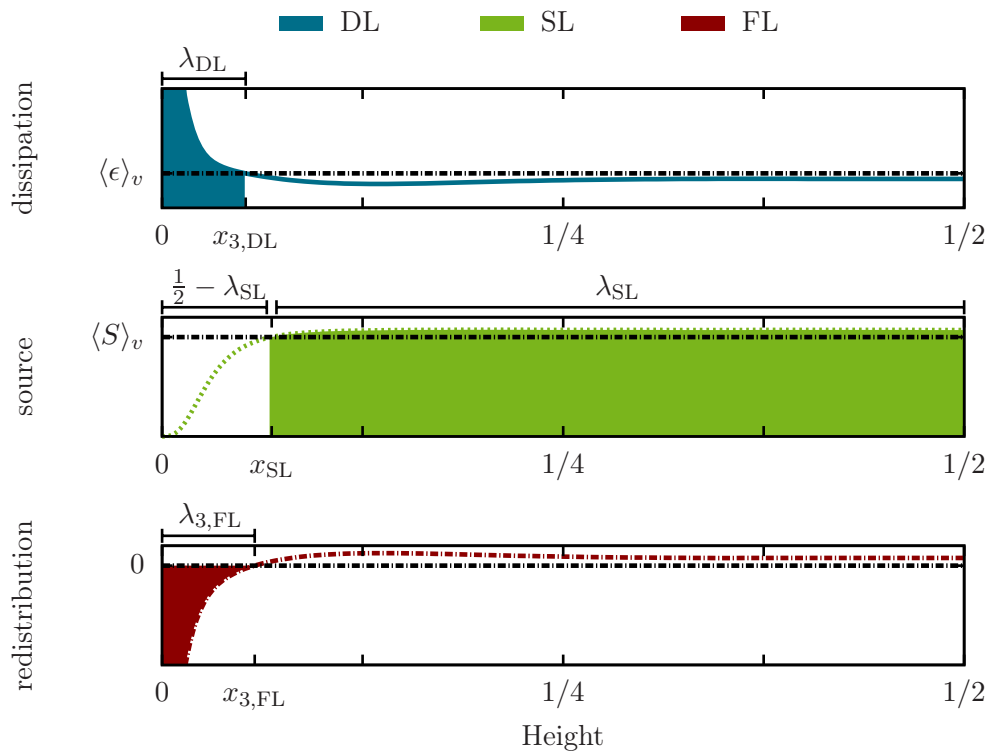}
     \caption{\label{fig:sketch-layer}
			Profiles of the dissipation, the source and the divergence of the flux along with the definition of the layers defined by the respective profile.}
\end{figure}
The observations on the profiles of the kinetic energy budget in the previous section (\S\ref{sec: profiles}) can be used to separate the domain into two regions, namely: a layer in which the profile is above the volume average and a layer in which the profile is below the volume average.
\citet{Petschel2013} have used the crossing between these two regions to define the so-called dissipation layer ($\mathrm{DL}$).
More specifically, the edge of the dissipation layer is defined by the vertical position at which the dissipation rate equals its volume-averaged value,
\begin{equation} 
     \left\langle \epsilon \right\rangle(x_{3,\mathrm{DL}})\stackrel{!}{=}\left\langle \epsilon\right\rangle_v\,.
\end{equation}
The distance from the closest boundary then defines the viscous dissipation layer thickness $\lambda_\mathrm{DL}$, as illustrated in figure \ref{fig:sketch-layer}.
The interpretation of the dissipation layer is rather simple: Fluid within the dissipation layer is characterised by a dissipation above the volume-average.

Similar to the dissipation layer, we introduce the source layer ($\mathrm{SL}$) which is characterised by enhanced kinetic energy production.
The edge of the source layer is defined analogous to the dissipation layer by the vertical position at which the kinetic energy source equals its volume-averaged value,
\begin{equation} 
     \left\langle S \right\rangle(x_{3,\mathrm{SL}})\stackrel{!}{=}\left\langle S \right\rangle_v. 
\end{equation}
Kinetic energy production due to buoyancy is an essential feature of the bulk region.
Consequently, we define the distance from the centre of the domain as the source layer thickness $\lambda_\mathrm{SL}$, c.f. figure~\ref{fig:sketch-layer}.
The dissipation and the source layer separate the flow into regions of high (above average) and low (below average) kinetic energy dissipation and production, respectively.

The kinetic energy flux transports kinetic energy from the source layer into the dissipation layer.
Thus we can define a flux layer ($\mathrm{FL}$) which separates the domain into regions of flux-induced reduction and accumulation of kinetic energy. 
The edge of the flux layer is then defined by the vertical position where the divergence of the flux equals zero, which corresponds to a local extremum of the flux, i.e.
\begin{equation} \label{eqn: flux layer}
    \pdiff{}{x_{3}} \left\langle J_{3} \right\rangle(x_{3,\mathrm{FL}})\stackrel{!}{=}0\quad\mbox{respectively}\quad\left\langle J_{3} \right\rangle(x_{3,\mathrm{FL}})\stackrel{!}{=}\mbox{max}\left|\left\langle J_{3} \right\rangle\right|\,.
\end{equation}
The distance from the closest boundary then sets the kinetic energy flux layer thickness $\lambda_\mathrm{FL}$.
From the definitions above it can be shown that the edge of the flux layer has to be located in between the edges of the dissipation and the source layer.
\begin{figure}
     \centering \includegraphics{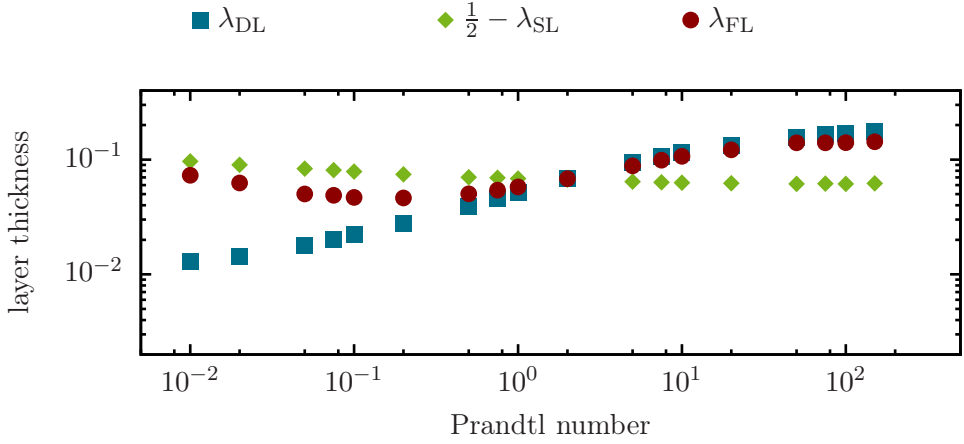}
     \caption{\label{fig:layer_pr_sink_source}
			Thickness of different layers versus the Prandtl number. Blue squares represent the thickness of the dissipation layer $\lambda_\mathrm{DL}$, red circles denote the flux layer thickness $\lambda_\mathrm{FL}$ and green diamonds indicate the distance between the boundary and the source layer $\thalf-\lambda_\mathrm{SL}$.}
\end{figure}
\subsection{Prandtl dependence of the dissipation, the source and the flux layer thicknesses}
Figure~\ref{fig:layer_pr_sink_source} shows the thickness of the dissipation layer $\lambda_\mathrm{DL}$ and of the flux layer $\lambda_\mathrm{FL}$ along with the distance between the boundary and the source layer ($\thalf-\lambda_\mathrm{SL}$) as a function of the Prandtl number.
The figure reveals that at low $\Pran$, the region of kinetic energy production and dissipation are spatially separated, while for high $\Pran\ge 2$, they overlap.
When they are spatially separated, the region in between is characterised by low production and dissipation of kinetic energy as can be seen by a closer inspection of the profiles (not shown here).
The overlap region, observed for high Prandtl numbers, features high energy production and dissipation levels.

The flux layer thickness displays no monotonic $\Pran$ dependence.
For low Prandtl numbers it compares approximately to the magnitude of the gap between the boundary and the source layer (i.e. $\lambda_\mathrm{FL}\approx\thalf-\lambda_\mathrm{SL}$).
For high Prandtl numbers it has roughly the same thickness as the dissipation layer.
 \begin{figure}
     \centering\includegraphics{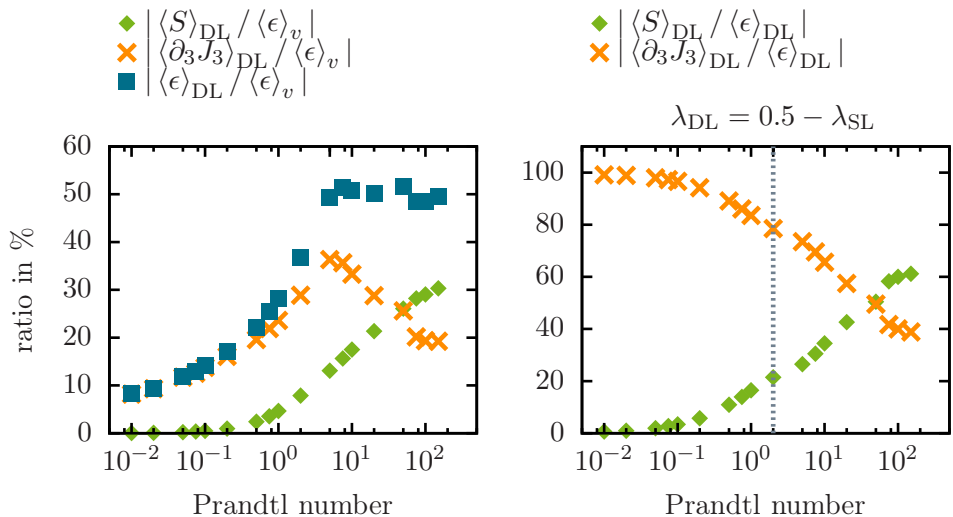}
     \caption{\label{fig: dissipation in dl}
			Ratio of dissipation layer contributions normalised by volume-averaged dissipation rate versus the Prandtl number (left graph). 
			Rate of kinetic energy accumulation in the dissipation layer normalised by the dissipation rate averaged over the dissipation layer (right graph). 
			The blue squares denote the dissipation in the dissipation layer, while the green diamonds denote the kinetic energy produced within the dissipation layer.
			Additionally orange crosses indicate accumulation of kinetic energy by the flux in the dissipation layer.
			The vertical grey dashed line indicates the Prandtl number where the source and the dissipation layers are of the same thickness.}
\end{figure}

\citet{Petschel2013} have argued for a transition from bulk-dominated dissipation at low $\Pran$ to significant dissipation within the dissipation layer at high $\Pran$.
As can be seen in the left graph of figure~\ref{fig: dissipation in dl}, high Prandtl numbers are characterised by a saturation of the relative dissipation within the dissipation layer.
Interestingly, the dissipation only saturates because the production increases; due to accumulation of kinetic energy by flux alone such a high dissipation in the dissipation layer is not possible.
Hence, the spatial overlap of production and dissipation has a strong influence on the net dissipation in the dissipation layer.
To examine the influence of this overlap closer, the right graph in figure~\ref{fig: dissipation in dl} shows the accumulation of kinetic energy in the dissipation layer by flux and production, the sum of which is balanced by dissipation.
As expected, the production within the dissipation layer vanishes for low Prandtl numbers because the source and the dissipation layer are spatially separated.
Consequently, the dissipation in the dissipation layer is balanced only by the accumulation of kinetic energy by flux for low Prandtl numbers.
For high Prandtl numbers, however, more kinetic energy is produced in the dissipation layer than is accumulated by flux.
This finding suggests that a dominance of the dissipation layer, as described by \citet{Petschel2013}, goes along with an overlap of the source and the dissipation layer.

\subsection{Flux layer thickness for individual fluxes\label{sec: layer thickness fluxes}}

\begin{figure}
     \centering \includegraphics{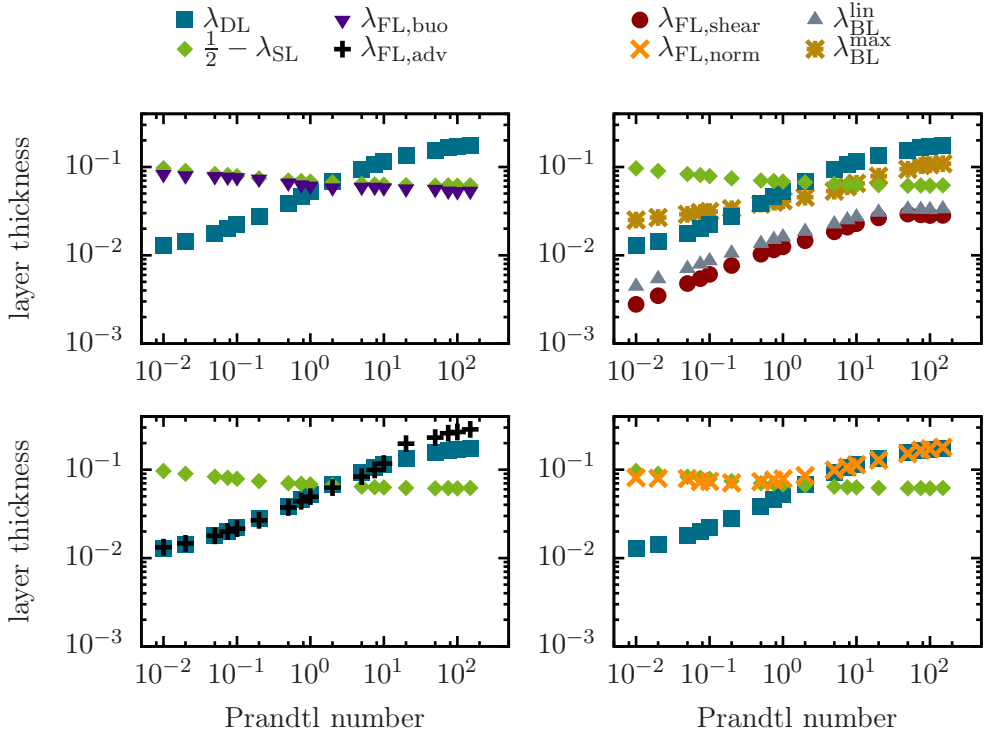}
     \caption{\label{fig: prandtl--blasius}
			Thicknesses of different layers versus the Prandtl number.
			The dissipation layer, represented by blue squares, and the source layer, represented by green diamonds, can be seen in all graphs.
			Additionally,  the flux layer thicknesses are denoted by purple triangles for the flux by buoyancy, black crosses for the flux by advection, orange crosses for flux by normal stresses and red dots indicate the shear flux layer thickness.
			In the upper right graph, grey triangles indicate a classical viscous boundary layer ($\lambda_\mathrm{BL}^{\mathrm{lin}}$), measured by the slope method, and golden stars indicate the position of the maximum of the horizontal velocity profile ($\lambda_\mathrm{BL}^{\mathrm{max}}$).}
\end{figure}

To understand the influence of the individual fluxes on the kinetic energy profiles more deeply, figure~\ref{fig: prandtl--blasius} shows the thickness of the four individual flux layers ($\lambda_\mathrm{FL, shear}$, $\lambda_\mathrm{FL, adv}$, $\lambda_\mathrm{FL, buo}$ and $\lambda_\mathrm{FL, norm}$) along with the dissipation layer ($\lambda_\mathrm{DL}$) and the distance from the boundary to the source layer ($\thalf-\lambda_\mathrm{SL}$).
Similar to the overall flux, the edge of the flux layers indicates the height of the transition from removal to accumulation by the corresponding flux (\ref{eqn: flux layer}).

As can be seen in the upper left graph of figure~\ref{fig: prandtl--blasius}, the thickness of the buoyancy-pressure flux layer varies only little with Prandtl number, and corresponds in magnitude to the gap between the source layer and the boundary.
The good agreement with the source layer suggests that buoyancy pressure is related to the kinetic energy production and therefore to the advective heat transport (cf.~\S\ref{sec: source profile}).
The advective flux layer (lower left graph) coincides with the dissipation layer for low $\Pran$, whereas the normal-stress flux layer (lower right graph) coincides with the dissipation layer for high $\Pran$.
This suggests that both fluxes -- the advective flux for low $\Pran$ and the normal-stress flux for high $\Pran$ -- transport kinetic energy into the dissipation layer.
For low Prandtl numbers the normal-stress flux layer equals the distance from the boundary to the source layer, indicating that normal stresses redistribute kinetic energy out of the source layer. 
A completely different picture can be observed for the shear-stress flux layer (upper right graph) which is much thinner than all other layers.
This suggests that shear stress redistributes kinetic energy only close to the walls, within the dissipation layer.

\subsection{Comparison with classical velocity boundary layers\label{sec: linear bl}}
In addition to the shear-stress flux layer, the upper right graph of figure~\ref{fig: prandtl--blasius} shows the thicknesses of two classical velocity boundary layers, namely: a linear boundary layer $\lambda_\mathrm{BL}^{\mathrm{lin}}$ and a maximum boundary layer $\lambda_\mathrm{BL}^{\mathrm{max}}$.
The thickness of the linear boundary layer $\lambda_\mathrm{BL}^{\mathrm{lin}}$ is defined by using linear fits of the horizontal velocity profiles in the vicinity of the boundary, i.e.~the so-called slope method \citep{Li2012,Scheel2012,Breuer2004}.
The linear boundary layer is often described as and compared to a Prandtl--Blasius boundary layer \citep{Zhou2011}.
The shear-stress flux layer largely coincides with the linear boundary layer $\lambda_\mathrm{BL}^{\mathrm{lin}}$ which agrees well with the view that a velocity boundary layer is characterised by a horizontal shear flow \citep{Schlichting2000}.

A further classical boundary layer definition, the maximum boundary layer $\lambda_\mathrm{BL}^{\mathrm{max}}$ \citep{Scheel2012,Breuer2004,Kerr2000}, can be related to the kinetic energy transport.
Since the thickness of the maximum boundary layer is defined by the position of the maximum of the root-mean squared velocity profile, the edge of the layer equals the height at which the shear stress flux profile becomes zero, cf. (\ref{eqn:av viscous shear flux}), and thus the transport direction of the flux reverses.
This means that the maximum boundary layer separates the domain into a region, $x_{3}\le\lambda_\mathrm{BL}^{\mathrm{max}}$, where shear stress transports kinetic energy towards the walls and into a region, $x_{3}\ge\lambda_\mathrm{BL}^{\mathrm{max}}$, where shear stress transports kinetic energy towards the center.
Consequently, shear stress flux cannot redistribute kinetic energy across the edge of the maximum boundary layer.
As the dissipation layer has approximately the same thickness as $\lambda_\mathrm{BL}^{\mathrm{max}}$ it thus follows that shear stress redistributes kinetic energy only within the dissipation layer.

The layer thicknesses displayed in figure~\ref{fig: prandtl--blasius} support the view that kinetic energy is transported mainly by advective or normal-stress fluxes from the bulk into the dissipation layer. 
Within the dissipation layer, shear stress redistributes kinetic energy towards the vicinity of the wall.
At the boundaries, where the strongest dissipation takes place, dissipation is balanced only by shear-stress fluxes.
This is in line with the picture of shear-flow driven velocity boundary layers.
The correlation of the shear flux layer and the linear boundary layer $\lambda_\mathrm{BL}^{\mathrm{lin}}$ also suggests that by focusing on a linear boundary layer one mainly measures kinetic energy redistributions by shear stress within the dissipation layer. 

\section{Kinetic Energy Redistributions\label{sec: deviations}}
Up to this point, we have discussed the profiles of the kinetic energy balance for different Prandtl numbers and have analysed the spatial separation of the domain by these profiles in dependence of the Prandtl number.
But so far, little has been said about the strength of kinetic energy redistribution as a function of the Prandtl number.
In this section, we first define a measure to investigate the strength of the redistribution of kinetic energy. 
The same measure can conveniently be used to assess the vertical inhomogeneity of the dissipation and production of kinetic energy.
We then analyse the strength of the redistributions and the vertical inhomogeneities in dependence of the Prandtl number.
In line with the previous sections, we investigate first the joint contributions and then discuss the individual fluxes.
\subsection{Quantification of the redistribution}
To quantify the redistribution of kinetic energy, we use the vertical standard deviation of the individual terms in the kinetic energy balance,
\begin{equation}\label{eqn:std dev}
     \sigma_{x_3}(X) =\sqrt{\left\langle\left(\left\langle X \right\rangle- \left\langle X \right\rangle_v \right)^2 \right\rangle_v} \,,
\end{equation}
where $\left\langle X \right\rangle$ is a profile and $\left\langle X \right\rangle_{v}$ its volume average.
Since for all flux terms the volume average is zero, the vertical standard deviation of the divergence of a flux ($\sigma_{x_3}(\partial_{3}J_{3})$) measures the root-mean-square (rms) redistribution of kinetic energy by each individual flux.
Consequently, in the following we use the term ``rms redistribution'' for the vertical standard deviation of the divergence of a flux term. For the source and dissipation, the vertical standard deviation provides a measure for the amount of vertical variation of the corresponding profile around the mean value. It can thus be interpreted as a measure for the vertical inhomogeneity of the source or dissipation profile. In the following, we thus refer to standard deviations of the source and dissipation profiles as the ``vertical source inhomogeneity'' and ``vertical dissipation inhomogeneity'', respectively. The consistent use of this terminology greatly simplifies the presentation of the results in the next section.
\subsection{Redistributions of kinetic energy in dependence of the Prandtl numbers}
 \begin{figure}
     \centering \includegraphics{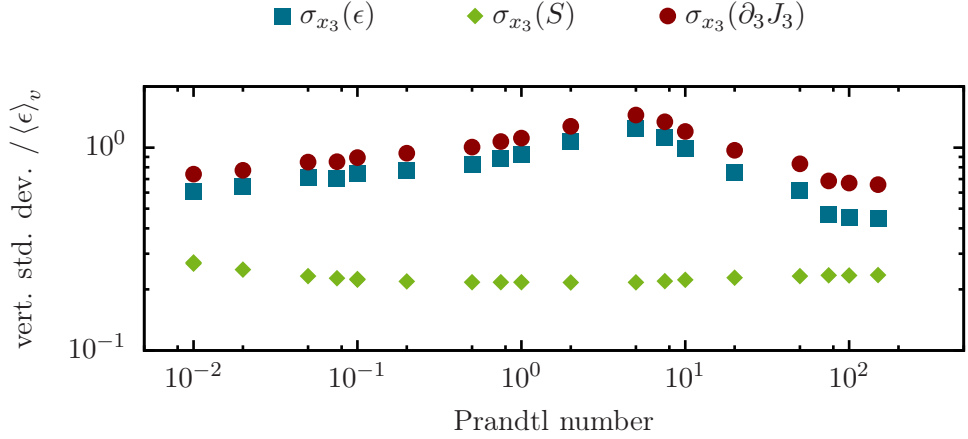}
     \caption{\label{fig: vertical_transport_rate}
			Vertical inhomogeneities of the profiles of dissipation, source and the redistributions by flux, measured by the vertical standard deviation of the respective profile, versus the Prandtl number.
			The blue squares denote the vertical dissipation inhomogeneity while the green diamonds denote the vertical source inhomogeneity and the red dots indicate the rms redistribution by flux.}
\end{figure}
A central result of the budget profiles in section \ref{sec: profiles} was that the source does not depend significantly on $\Pran$ while the kinetic energy sink and flux do.
This observation can be seen more quantitatively in figure~\ref{fig: vertical_transport_rate} which shows that the vertical source inhomogeneity is small for all $\Pran$, whereas the vertical dissipation inhomogeneity and the redistributions varies as a function of the Prandtl number.
It is observed that, apart from a constant prefactor in amplitude, the rms redistributions show a similar Prandtl number dependence as the vertical dissipation inhomogeneity: Both quantities increase with $\Pran$ for low Prandtl numbers, form a maximum at $\Pran\approx5$ and then decrease again for high Prandtl numbers.
These findings reveal that redistributions are related to inhomogeneities of the dissipation.

The non-monotonic $\Pran$ dependence of the dissipation with a maximum at moderate Prandtl numbers suggests a competition of two effects:
The decreasing transport of kinetic energy with decreasing $\Pran$ for low Prandtl numbers might be the consequence of a homogenisation due to increased bulk turbulence.
This seems plausible in view of the fact that the Reynolds number increases for decreasing Prandtl numbers, and is consistent with a bulk-dominated flow, as predicted by \citet{Grossmann2000}.
The decreasing transport for high Prandtl numbers is likely to be caused by a smoothing of the velocity field due to diffusion at high $\Pran$, which leads to more homogeneous dissipation profiles.
This process becomes dominant around $\Pran\approx5$, and obviously outweighs the effect of enhanced dissipation near the walls \citep{Grossmann2000}.

\subsection{Redistributions by individual fluxes}
\begin{figure}\centering
     \includegraphics{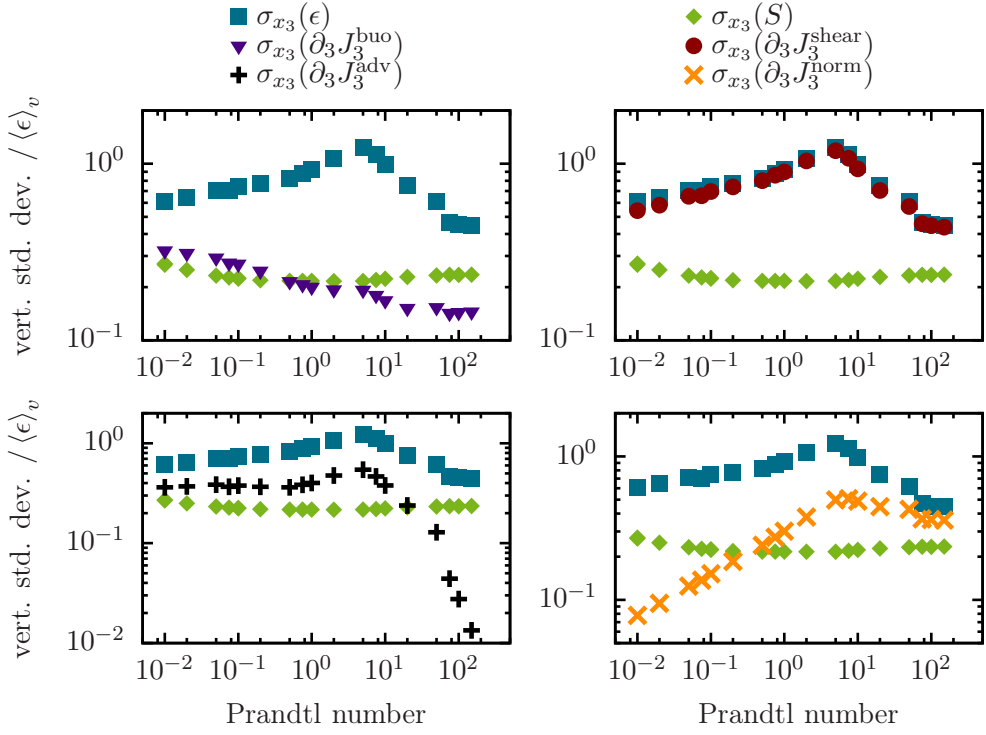}
     \caption{\label{fig: vertical_transport_rate_decomp}
			Root-mean-square redistributions by different fluxes along with the vertical source and dissipation inhomogeneities, measured by the vertical standard deviation of the respective profile, versus the Prandtl number.
			The blue squares denote the vertical dissipation inhomogeneity while the green diamonds denote the vertical source inhomogeneity.
			In the upper left graph the redistribution by buoyancy-pressure flux is indicated by purple triangles, red dots indicate redistribution by viscous shear flux in the upper right graph, black crosses indicate redistribution by advective flux in the lower left graph and orange crosses denote redistributions by the viscous normal flux in the lower right graph.}
\end{figure}
For a more detailed analysis of the Prandtl number dependence of the dissipation, figure~\ref{fig: vertical_transport_rate_decomp} shows the rms redistributions by the individual fluxes. 
It is observed that, apart from the buoyancy-pressure flux, the rms redistributions by all fluxes separate the observed parameter space into two regimes.
The transition between these regimes takes place at $\Pran\approx5$ where the vertical dissipation inhomogeneity also exhibits a maximum.

Similar to the layer thicknesses, the redistribution by buoyancy pressure (see figure~\ref{fig: vertical_transport_rate_decomp} top left) is related to the kinetic energy source.
Both, the amplitudes as well as the Prandtl number dependence, are comparable to the vertical source inhomogeneity.
The rms redistribution decreases for increasing $\Pran$ throughout the parameter range covered by our study.

A common view of describing convection for different Prandtl numbers is to assume that advection plays a major role for low Prandtl numbers whereas high Prandtl numbers are characterised by viscous effects \citep{Grossmann2000,Breuer2004}.
The redistributions by advection (figure~\ref{fig: vertical_transport_rate_decomp} bottom left) and by normal stress (figure~\ref{fig: vertical_transport_rate_decomp} bottom right) support these assumptions.
In the low Prandtl number regime the normal-stress redistributions decrease monotonically with decreasing $\Pran$, whereas the advective redistributions saturate for small $\Pran$. 
In the high Prandtl number regime the normal-stress redistributions saturate at high values whereas the advective redistributions decrease rapidly with increasing $\Pran$.
The dominance of viscous effects for high $\Pran$ is often related to a growing velocity boundary layer \citep{Grossmann2000}.
In this case, however, the kinetic energy transport by viscous effects can be related to normal stresses and not to shear stresses and is therefore not associated with boundary layer processes. 

\begin{figure}\centering
     \includegraphics{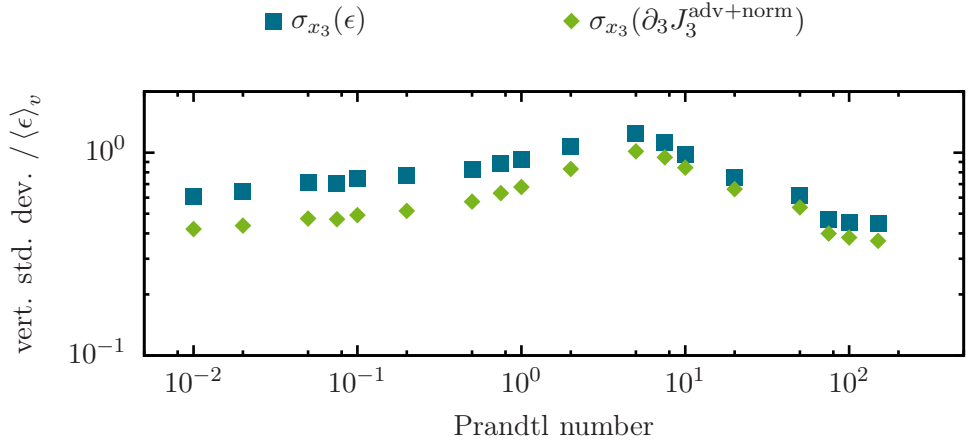}
     \caption{\label{fig: vertical_transport_rate_sum}
			Root-mean-square redistribution of kinetic energy by the sum of advective flux and normal-stress flux along with the vertical dissipation inhomogeneity, measured by the vertical standard deviation of the respective profile, versus the Prandtl number.
			The blue squares denote the vertical inhomogeneities of the dissipation profile while the green diamonds denote redistributions by the flux, respectively.}
\end{figure}
Classical boundary layer processes only become dominant closer to the walls. In fact, it is observed that the shear-stress flux layer is about the size of the velocity boundary layer (cf. figure~\ref{fig: prandtl--blasius}). 
The rms redistributions by shear stress (figure~\ref{fig: vertical_transport_rate_decomp} top right) display a nearly identical Prandtl number dependence to the vertical dissipation inhomogeneity.
These similarities may be related to the balance of both quantities at the boundaries.
The identical Prandtl number dependence of the redistributions by shear and the vertical dissipation inhomogeneity might naively be interpreted as a strong influence of the shear stress on the kinetic energy dissipation profile.
But the redistribution by shear stress takes place much closer to the wall than all other redistributions (cf. figure~\ref{fig: prandtl--blasius}). 
This points out that advection and normal stresses redistribute kinetic energy from the source layer into the dissipation layer, from where it is further redistributed into the boundary layer by shear stresses.
Because the shear-stress redistributions follow the vertical dissipation inhomogeneity for all Prandtl numbers, shear stress cannot be associated with the decreasing vertical dissipation inhomogeneity in only one of the two $\Pran$ regimes.

The opposing Prandtl number dependence of the advective and normal-stress redistributions indicates two competing Prandtl number dependencies which may result in a maximum at moderate $\Pran$.
This becomes more clear by analysing the redistributions by the sum of both fluxes which can be seen in figure~\ref{fig: vertical_transport_rate_sum} along with the vertical kinetic energy dissipation inhomogeneity.
It can be seen that the redistribution by the sum of both fluxes indeed displays a Prandtl number dependence similar to the vertical inhomogeneities of the dissipation profile.
Only for low Prandtl numbers an approximately constant prefactor in amplitude distinguishes the redistribution and the vertical dissipation inhomogeneity, which appears to be caused by the increasing influence of the kinetic energy redistributions by buoyancy pressure.
In summary, in order to explain the dissipation profile, both shear-stress flux close to the boundary and the combined effect of advective and normal-stress flux need to be considered.

\section{Summary and Discussion\label{sec: discussion}}
In this work we have extensively analysed the contributions to the kinetic energy budget  in Rayleigh--B\'{e}nard convection for a range of Prandtl numbers at a fixed Rayleigh number. 
The presented results allow to draw a comprehensive picture of kinetic energy transport. 
On a very general level, we observe that the kinetic energy is produced in the central region where the temperature field is well mixed, from where it is transported outwards and dissipated near the walls. 
This is consistent with the observations of earlier studies by \citet{Deardorff1967} and \citet{Kerr2001}. 
It is also intuitive because fluid is accelerated due to buoyancy forces in the bulk, and decelerated when approaching the boundaries of the volume.

\begin{figure}
     \centering \includegraphics{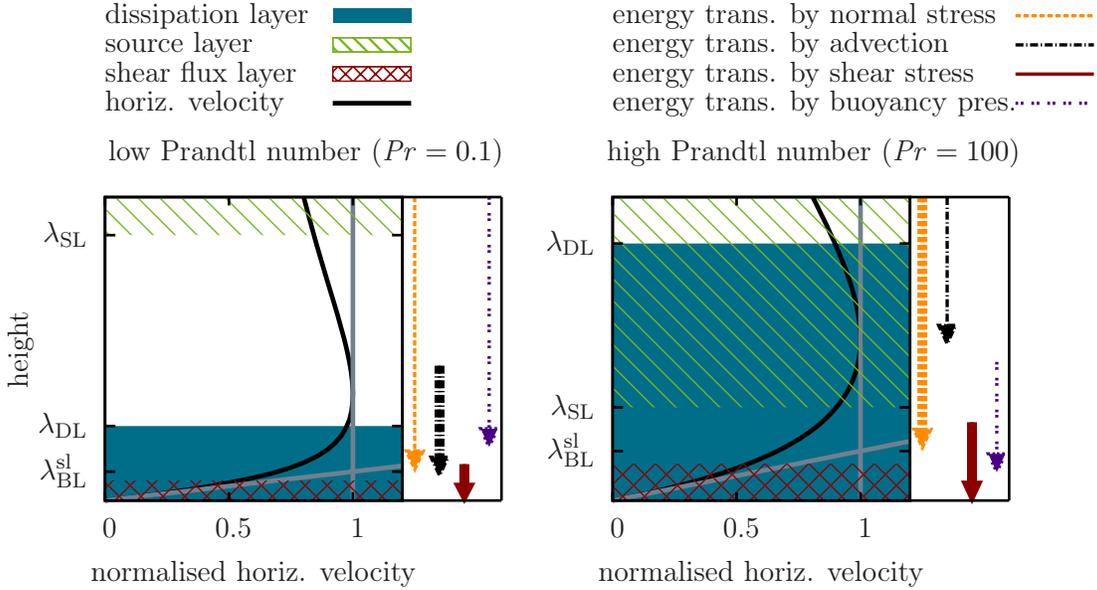}\caption{\label{fig: sketch}
Sketch of the different layers of the kinetic energy balance along with the profile of the rms of horizontal velocity normalised by its maximum for $\Pran=0.1$ and $\Pran=100$.
Additionally the spatial redistribution of kinetic energy is represented schematically by the arrows on the right margin.
The starting point of these arrows indicate the height of maximal removal of kinetic energy whereas the endpoint indicate the height of maximal accumulation of kinetic energy.
The broad arrows denote the dominant flux processes.
}
\end{figure}

On a more detailed level, our work provides a comprehensive view of the physical processes involved in transporting the kinetic energy from the bulk towards the walls.
The dynamics is found to be surprisingly rich and exhibits a pronounced Prandtl number dependence.
The investigated Prandtl-number range can be separated into two characteristic regimes, corresponding to high and low Prandtl numbers. 
Figure~\ref{fig: sketch} shows a schematic sketch of the near-wall region for two exemplary Prandtl numbers ($\Pran=0.1$ and $\Pran=100$). 
Shown are regions of enhanced dissipation (blue solid areas) and of enhanced production (green shaded areas) of kinetic energy. Arrows indicate the kinetic energy fluxes, each of which is caused by a different process. The arrows point from the height of maximal removal to the height of maximal accumulation of kinetic energy, with bold lines indicating dominant contributions. 

Perhaps the most striking feature is that the dissipation and the source regions are spatially separated for low Prandtl numbers, whereas they penetrate each other for high Prandtl numbers. This spatial overlap of source and dissipation region is found to be crucial in allowing the boundary region to dominate the overall dissipation at high $\Pran$ \citep{Grossmann2000,Petschel2013}. The kinetic energy fluxes alone are too weak to balance the dissipative kinetic energy losses within this region. 
For low Prandtl numbers, the source and dissipation regions do not overlap, such that the dissipative losses have to be balanced almost entirely by fluxes transporting kinetic energy over the spatial gap between the source and dissipation region. These fluxes are mainly accomplished by advection.
A strong kinetic energy flux from the bulk to the boundary region is also observed at large $\Pran$. Instead of advection, normal viscous stresses dominate the transport in this case.  
Interestingly, for all Prandtl numbers, viscous shear stresses merely redistribute kinetic energy within the dissipative region and play a crucial role in balancing the viscous dissipation on the rigid walls. 

In order to put these results on a more quantitative footing, we proposed exact layer definitions reflecting the above picture. Extending the concept of dissipation layers, introduced earlier by \citet{Petschel2013}, to source layers and flux layers, our work suggests a decomposition of the convective domain into several distinct sublayers, each characterised by different physical processes at work, either producing, dissipating or transporting kinetic energy. These new layers prove to be useful in contextualising previous work relying on specific definitions of mechanical boundary layers. 

Popular boundary layer definitions include the so-called slope-method or are based on the maximum of the horizontal velocity profile. While originally introduced heuristically, our work reveals that they are deeply connected to the kinetic energy transport. The edge of the maximum boundary layer is reinterpreted as the location from where on viscous shear stresses begin to transport kinetic energy towards the boundary. In contrast, the edge of a boundary layer defined by the slope method is found to mark the location where viscous shear stresses begin to accumulate kinetic energy close to the wall. Both definitions however fail to correctly represent the region of enhanced dissipation, in contrast to the dissipation layer introduced by \citet{Petschel2013}. This is in line with recent results by \citet{Scheel2014}, who show that boundary layers determined by the slope method and dissipation layers exhibit opposite Rayleigh number scalings.

For the current work we have considered a wide range of Prandtl numbers for a fixed, moderate Rayleigh number. 
In the future, it would also be very interesting to see how our observations vary when the Rayleigh number is increased.
\bibliographystyle{jfm}

\bibliography{Library}

\end{document}